\documentclass{article}
\usepackage{ijcai24}

\usepackage{times}
\usepackage{soul}
\usepackage{url}
\usepackage[hidelinks]{hyperref}
\usepackage[utf8]{inputenc}
\usepackage[small]{caption}
\usepackage{graphicx}
\usepackage{amsmath}
\usepackage{amsthm}
\usepackage{booktabs}
\usepackage{algorithm}
\usepackage{algorithmic}
\usepackage[switch]{lineno}

\usepackage{newfloat}
\usepackage{listings}

\usepackage{pifont}
\usepackage{amssymb}
\usepackage{enumitem}

\usepackage{xcolor}
\usepackage{multirow}
\usepackage{array}
\usepackage{subcaption}
\newcolumntype{j}{>{\centering\arraybackslash}m{4.5cm}}
\newcolumntype{k}{>{\arraybackslash}m{9.5cm}}
\newcolumntype{K}{>{\centering\arraybackslash}m{2cm}}
\newcolumntype{a}{>{\arraybackslash}m{5.4cm}}
\newcolumntype{b}{>{\arraybackslash}m{3.53cm}}
\newcolumntype{A}{>{\arraybackslash}m{1.33cm}}
\newcolumntype{J}{>{\centering\arraybackslash}m{13.5cm}}

\title{Media Bias Matters: Understanding the Impact of Politically Biased News\\ on Vaccine Attitudes in Social Media}

\author{
Bohan Jiang$^1$
\and
Lu Cheng$^2$\and
Zhen Tan$^{1}$\and
Ruocheng Guo$^3$\And
Huan Liu$^{1}$\\
\affiliations
$^1$School of Computing and AI, Arizona State University\\
$^2$Department of Computer Science, University of Illinois Chicago\\
$^3$ByteDance Research\\
\emails
\{bjiang14, ztan36, huanliu\}@asu.edu,
lucheng@uic.edu,
ruocheng.guo@bytedance.com
}

\begin{document}

\maketitle

\begin{abstract}
News media has been utilized as a political tool to stray from facts, presenting biased claims without evidence. Amid the COVID-19 pandemic, politically biased news (PBN) has significantly undermined public trust in vaccines, despite strong medical evidence supporting their efficacy. In this paper, we analyze: (\textit{i}) how inherent vaccine stances subtly influence individuals' selection of news sources and participation in social media discussions; and (\textit{ii}) the impact of exposure to PBN on users' attitudes toward vaccines. In doing so, we first curate a comprehensive dataset that connects PBN with related social media discourse. Utilizing advanced deep learning and causal inference techniques, we reveal distinct user behaviors between social media groups with various vaccine stances. Moreover, we observe that individuals with moderate stances, particularly the vaccine-hesitant majority, are more vulnerable to the influence of PBN compared to those with extreme views. Our findings provide critical insights to foster this line of research.

\end{abstract}

\section{Introduction}

The pervasive impact of the COVID-19 pandemic transcends geographic and social boundaries~\cite{haleem2020effects}, having claimed over 7 million lives globally as of February 2024\footnote{\url{https://covid19.who.int/}}. Although vaccination emerges as the most efficacious defense, a substantial proportion of the population has shown \textit{vaccine hesitancy}~\cite{dror2020vaccine}. A KFF survey~\cite{kirzinger2021kff} reported that more than 40\% of parents in the U.S. are hesitant to get their children vaccinated due to safety and efficacy concerns. Meanwhile, hundreds of mainstream media are responsible for publishing COVID-related \textit{politically biased news} (PBN), turning the vaccine campaign into a political campaign~\cite{bolsen2022politicization}. While the skepticism around vaccines isn't new, when intertwined with news media, it exacerbates polarized opinions and conspiracy theories, thereby posing substantial challenges to public health efforts~\cite{sorell2022politics}.


\begin{figure}[!t]
\centering\scalebox{1}{
\includegraphics[width=1\columnwidth]{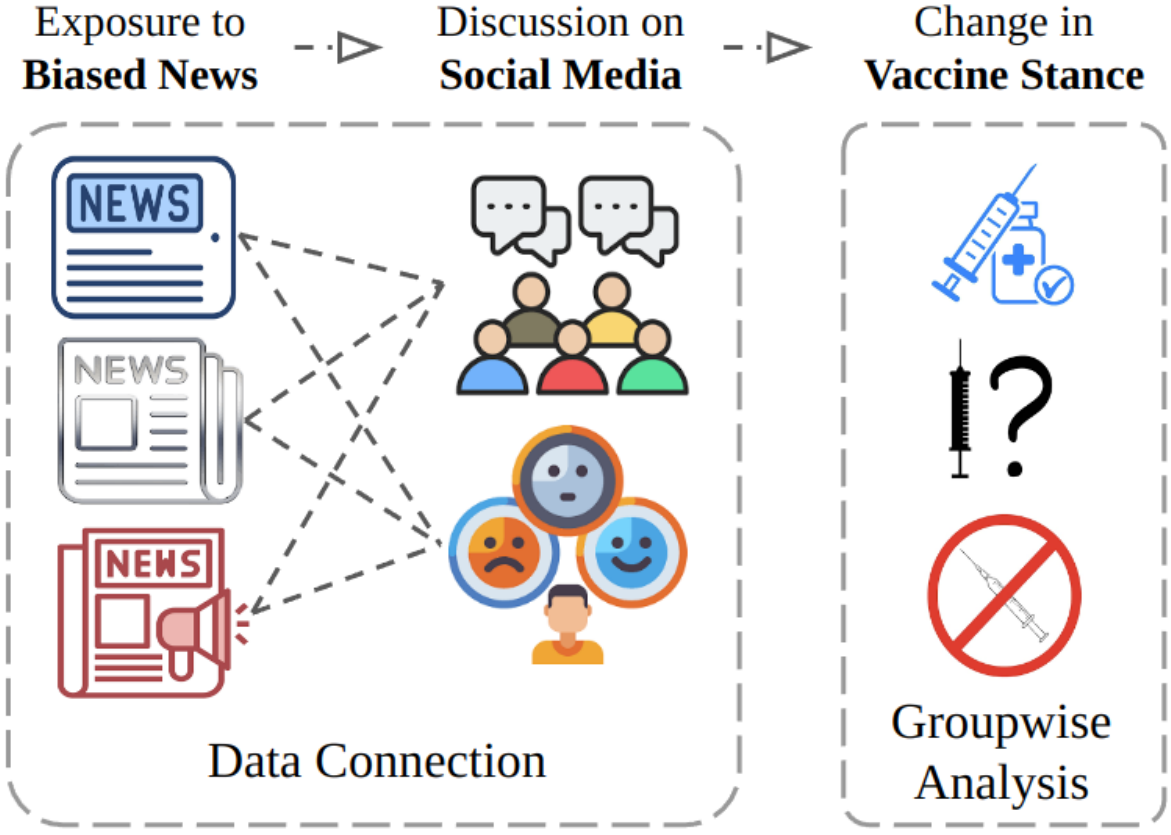}}
\caption{An overview of our research pipeline. On the left, it depicts news outlets disseminating COVID-related PBN to wide audiences via social media platforms. The right side illustrates the potential influence of such PBN exposure on users' vaccine stances. We define \textit{exposure} as instances where a user engages with PBN through Retweets or Quote-tweets (Retweets with added comments). \textit{Discussion} refers to user involvement in vaccine-related conversations after PBN \textit{exposure}. \textit{Change} denotes the variation in users' vaccine stances resulting from PBN \textit{exposure}.}
\label{pipeline}
\end{figure}



    
    

Recent research has examined the impact of news coverage~\cite{le2022understanding}, 
political polarization~\cite{ebeling2022analysis}, and misinformation dissemination in social media~\cite{miyazaki2023fake}.
However, few efforts have been focused on: (\textit{i}) providing a comprehensive dataset that connects news media data with associated social media data; and (\textit{ii}) understanding the causal relationship between exposure to PBN and people's vaccine stance changes. Most existing works~\cite{joseph2022local,poddar2022winds,spiteri2021media}
studied their \textit{correlation} instead of \textit{causation}, the latter of which is the key to understanding the impact of a PBN intervention (i.e., reading a COVID-vaccine-related PBN) on the outcome (i.e., COVID-19 vaccine stance changes).
While \cite{fowler2022effects} examined the causal relationship via surveys, it is limited to a small sample size and overlooks potential confounders such as user heterogeneity and social media features related to treatment and outcome. Modeling confounders is challenging due to the scarcity of observational data and domain knowledge. Moreover, little existing work differentiates the causal effect of groups with various vaccine stances. Large-scale studies on social media can complement prior research in the field to better understand how PBN shapes people's stances toward vaccines. 

To address aforementioned limitations and challenges, we construct \textbf{CovNS} (\textit{\textbf{\underline{Cov}}id-19 biased \textbf{\underline{N}}ews and \textbf{\underline{S}}ocial media dataset}), the first dataset which bridges the gap between COVID-related PBN and the corresponding social media discussions. Utilizing \textbf{CovNS}, we propose a research pipeline (see Figure~\ref{pipeline}) to study the following research questions:
\begin{itemize}[leftmargin=*]
    \item \textbf{RQ1:} \textit{How do PBN consumption and social media discussion vary with different vaccine stance groups?} 
    \item \textbf{RQ2:} \textit{To what extent does PBN exposure contribute to the reversal and reinforcement of vaccine stances?}
\end{itemize}

In summary, this study makes the following contributions to systematically address the posed research questions:
\begin{itemize}[leftmargin=*]
    \item \underline{\textit{Dataset Curation}}: We build \textbf{CovNS}, the first dataset that establishes a connection between the COVID-related PBN and the social media data, enabling future research by providing sufficient annotations.
    \item \underline{\textit{Behavior Analysis}}: We characterize pro-vaccine, anti-vaccine, and vaccine-hesitant groups on Twitter. We delve into the differences in PBN consumption preference and social media discussions across these groups.
    \item \underline{\textit{Causal Effect Estimation}}: We employ advanced causal learning methods to estimate the causal effect of reading PBN on users’ vaccine stance changes.
\end{itemize}

Our observations reveal that the three vaccine stance groups demonstrate significantly \textit{different} user behaviors. Meanwhile, exposure to left-leaning and right-leaning news sources generally causes social media users to be more pro-vaccine and anti-vaccine, respectively. Notably, users who are hesitant about vaccination show a greater \textit{vulnerability} to the influence of PBN, underscoring the significant challenges faced in promoting vaccination campaigns.

\section{Related Work}
In this section, we review three existing bodies of literature that lay the foundation for this work: (\textit{i}) long-standing literature on the political bias in news media in the United States; (\textit{ii}) fast-rising literature on the COVID-19 vaccine stance; and (\textit{iii}) well-established work on applying causal inference methods on COVID-related problems.

\subsection{Political Bias in News Media}
Political bias in news media has been studied extensively in the areas of political science, social science, etc~\cite{eberl2017one}.
It is an inherent bias of journalists and media outlets that makes them intentionally report biased news articles in order to serve a political agenda. The alteration of the news content usually operates in two ways: (i) \textit{issue framing}, i.e., presenting an issue in a way that will likely get the most agreement from supporters; and (ii) \textit{issue filtering}, i.e., selectively omit information that supports an alternative opinion on the other political side~\cite{iyengar1994anyone}.
As a result, readers were manipulated by misleading or false viewpoints and narratives. Prior work examining political bias using both qualitative and quantitative methods, has shown that U.S. news media differs ideologically and can create a highly polarized social environment~\cite{budak2016fair}. During the COVID-19 pandemic, much scholarly attention has been devoted to understanding the influence of PBN on public health. For example, some have shown that misleading information downplays the severity of COVID~\cite{teng2022characterizing}, false claims prevent people from knowing the fact~\cite{seo2022if}, and conspiracy communities use PBN to distance users from science~\cite{sharma2022covid}. 

\subsection{COVID-19 Vaccine Stance}
Various studies have investigated the COVID-19 vaccine stance in social media. Among them, some works contributed to providing labeled datasets regarding COVID-19 vaccine stance~\cite{mu2023vaxxhesitancy,glandt2021stance}
These works use either manual or algorithmic annotations to label a post as \textit{anti-vaccine}, \textit{pro-vaccine}, or \textit{vaccine-hesitant}. There exist other works that focus on building machine learning models to predict the COVID-19 vaccine stance, using linguistic features~\cite{poddar2022winds}, auxiliary information~\cite{tahir2022improving}, or large language models~\cite{riaz2022tm}. Besides, another line of research has been devoted to collecting COVID-vaccine-related datasets from news media.~\cite{semeraro2022writing} collected 5,745 news from 17 Italian news media.
\cite{joseph2022local} collected 750k articles from over 300 local news outlets to analyze relations between news coverage and offline behaviors. 

However, little prior work provided labeled datasets of COVID-vaccine-related PBN and social media discussion. To bridge this gap, our dataset combines both social media (e.g., tweets) and news media (e.g., news articles) data with multi-level manual and algorithmic annotations on news articles, posts, and users.

\subsection{Causal Analysis of COVID-Related Factors}
Well-established literature is built around learning causality with machine learning and big data~\cite{guo2020survey}. Much of the recent work examined the causal relationships among COVID-related factors. For instance,~\cite{hsiang2020effect} and~\cite{ma2022assessing} estimated the causal effect of different government efforts (e.g., COVID-19 policies) on offline statistics (e.g., number of infections and deaths). Other studies investigated the causal impact of online COVID-19 misinformation on one's mental health~\cite{verma2022examining} or vaccine hesitancy~\cite{pierri2022online}. \cite{fowler2022effects} conducted a survey to study the effect of exposure to politicized media coverage on people's negative emotional responses.

However, there is little understanding of the impact of consuming COVID-related PBN on social media users' willingness toward the COVID-19 vaccine uptake. As a remedy, we work on this problem by exploiting advanced causal machine-learning methods on the collected real-world dataset and proposed causal graphs.

\section{Bridging News Media and Social Media}
\begin{figure}[!t]
\centering\scalebox{1}{
\includegraphics[width=1\columnwidth]{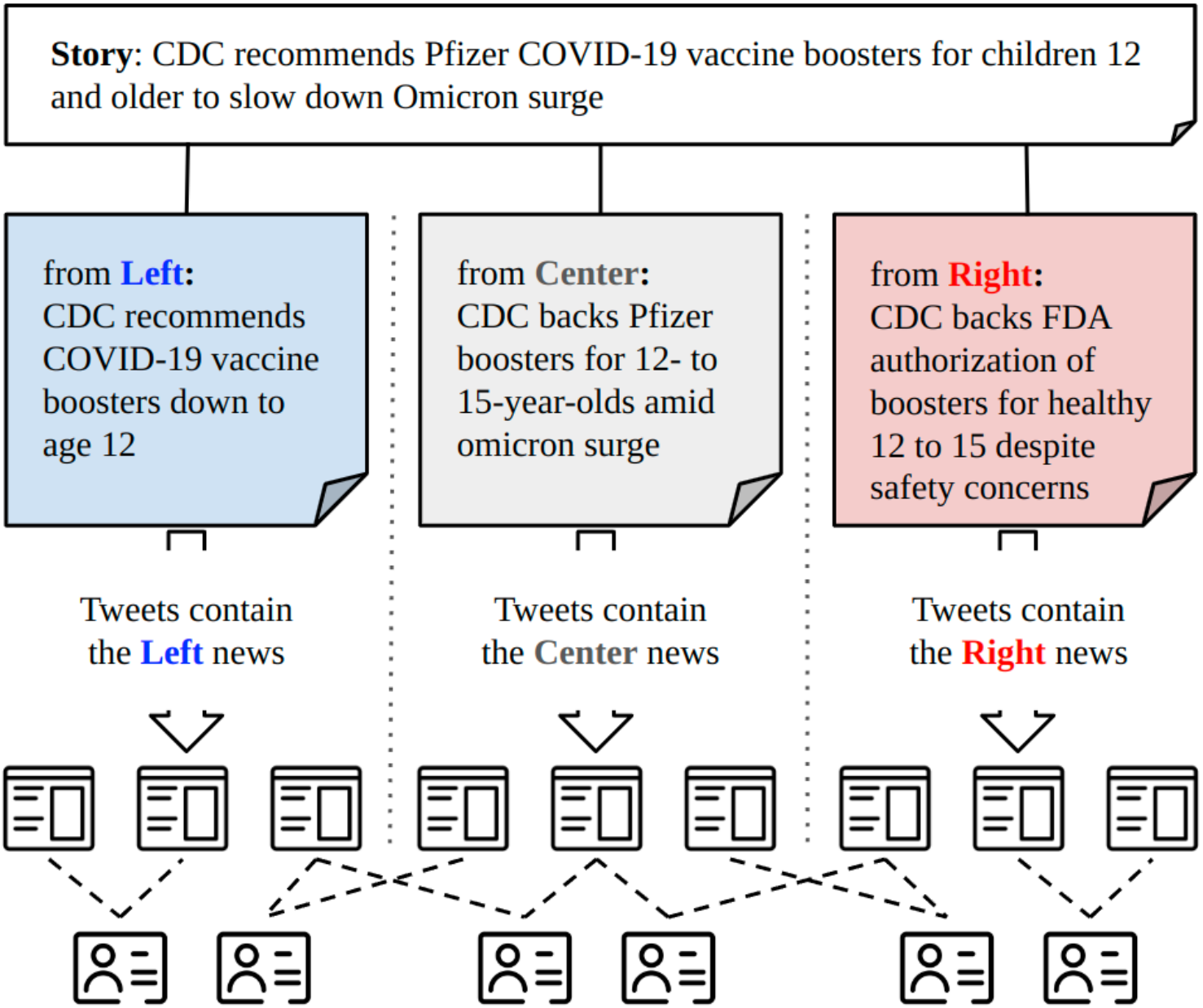}}
\vspace{-.5cm}
\caption{An example of the data collection process. We first collect a set of COVID-related news triplets containing articles from left-leaning, center-leaning, and right-leaning outlets from \textit{Allsides} (top). We then obtain associated \textit{Twitter} data (bottom).}
\vspace{-.1cm}
\label{dataset}
\end{figure}

In this section, we introduce how we build the dataset, including detailed data collection, annotation, and selection process. Figure~\ref{dataset} exemplifies the data collection process. We rely on two resources: \textit{Allsides}\footnote{\url{https://www.allsides.com/unbiased-balanced-news}} and \textit{Twitter}, to collect PBN and social media data, respectively. We obtain human-annotated labels such as political leaning from \textit{Allsides}. We then annotate the vaccine stances for the collected data using manual and algorithmic labeling strategies. At last, we apply filters to compose a high-quality dataset for our experiment. Table~\ref{tab:datasetoverview} provides an overview of \textbf{CovNS}.

\begin{table}[!t]
    \centering
    \resizebox{\linewidth}{!}{\small
    \begin{tabular}{lbcc}
        \toprule
         & \text{Fields} & \text{Stats} & \text{Labels}\\
        \cmidrule{1-4}
        \multirow{2}{*}{\textbf{\textit{Twitter}}} & \small Tweets & \small 243,412,961 & \small VS \\ 
        & \small User metadata (e.g., id) & \small 36,172 & \small VS \\
        \cmidrule{1-4}
        \multirow{4}{*}{\textbf{\textit{Allsides}}} & \small Stories (news triples) & \small 732 & \small VS \\    
        & \small News articles & \small 2,196 & \small VS, PL \\
        & \small News metadata (e.g., url) & \small 2,196 & - \\
        & \small News Media & \small 160 & \small PL \\
        \bottomrule
    \end{tabular}}
    \vspace{-.2cm}
    \caption{Overview of \textbf{CovNS}. Note that VS = vaccine stance; PL = political leaning.}
    \vspace{-0.2cm}
    \label{tab:datasetoverview}
\end{table}

\subsection{Data Collection}
\paragraph{PBN from Allsides.} 
We gather COVID-related PBN from \textit{Allsides}, 
a website that assesses the political bias of prominent media outlets, and presents different versions of similar news stories from sources of the political right, left, and center. It shows readers news coverage and diverse perspectives from all across the American political spectrum. In addition, it also provides a neutral summary to recapitulate the story and discuss how different news outlets spin or manipulate the facts. Note that \textit{Allsides} currently only focuses on English news from American media outlets.

For each news story, we collect the triplets of PBN~$\{\mathcal{N}_{L}, \mathcal{N}_{C}, \mathcal{N}_{R}\}$ denoting news articles from left-leaning, center-leaning, and right-leaning news media, respectively. They are associated with corresponding
titles $\{\mathcal{T}_{L}, \mathcal{T}_{C}, \mathcal{T}_{R}\}$, contents~$\{\mathcal{C}_{L}, \mathcal{C}_{C}, \mathcal{C}_{R}\}$, and URLs~$\{\mathcal{U}_{L}, \mathcal{U}_{C}, \mathcal{U}_{R}\}$. We also collect meta-information such as publication date, topics, media name, and URLs of banners or pictures for each news article. In total, there are 732 COVID-related news triplets (i.e., 2,196 news articles) from 160 U.S. media outlets.

\paragraph{Social Media Data from Twitter.}
We construct a large-scale Twitter dataset using the official academic research API. Following previous work on collecting news articles in social media~\cite{shu2020fakenewsnet}, we use the \underline{URL} of original PBN as the \underline{search query} on Twitter to collect all the available social media discourses such as tweets, retweets, and replies. Moreover, we collect historical tweets and profile information from users who have participated in the discussion about COVID-related PBN. Overall, this dataset consists of 243,412,961 historical tweets from 36,172 unique accounts.

\subsection{Data Annotation}
Two sets of labels are essential to analyze the relationship between PBN consumption and vaccine stance changes of social media users: \underline{political leaning} and \underline{vaccine stance}.


We adopt labels on political leaning (\textit{left}, \textit{lean-left}, \textit{center}, \textit{lean-right}, or \textit{right}) from \textit{Allsides}. For each news article, the rating process includes (\textit{i}) editorial review, (\textit{ii}) blind bias survey, (\textit{iii}) independent review, (\textit{iv}) third-party review, (\textit{v}) community feedback, and (\textit{vi}) confidence level.\footnote{\url{https://www.allsides.com/media-bias/media-bias-rating-methods}} However, both data sources (\textit{Allsides} and \textit{Twitter}) do not provide human-annotated labels on COVID-19 vaccine stances. As it is a time-consuming and labor-intensive task to annotate vaccine stances for the entire collected dataset with 2,196 news articles and 243,412,961 tweets, we follow the method mentioned in \cite{lyu2022social} and leverage a human-in-the-loop machine-learning strategy to minimize the manual annotation effort meanwhile maintaining high-quality labels. 

Specifically, we first identify common COVID-related keywords (\textit{covid}, \textit{coronavirus}, and \textit{SARS-CoV-2}) to filter out news articles and tweets unrelated to COVID-19. After that, we invite two annotators in the area to inspect the headline and content to assign one of the four labels for each news article:
\begin{itemize}[leftmargin=*]
    \item \textit{Pro-vaccine}, i.e., news that promotes the willingness of vaccine acceptance and uptake;
    \item \textit{Anti-vaccine}, i.e., news that discourages the willingness of vaccine acceptance and uptake;
    \item \textit{Mixed}, i.e., news that contains controversial opinions about COVID-19 vaccines; and
    \item \textit{Other}, i.e., news that contains general COVID-19 information, but are unrelated to vaccines.
\end{itemize}
Before annotating, we inform the annotators of a few examples of each category. With the collected annotations, we calculate Cohen's Kappa Score ($k$) to assess the inter-annotator agreement for the selected PBN dataset. We get $k$ = 0.83, which is considered ``almost perfect'' agreement according to~\cite{cohen1960coefficient}. To further obtain the ground truth, we only considered the annotations that both annotators agreed on. This results in 410 \textit{pro-vaccine}, 395 \textit{anti-vaccine}, 409 \textit{mixed}, and 772 \textit{other} news articles.

For our large-scale \textit{Twitter} dataset, we finetune a pre-trained CT-BERT~\cite{muller2020covid} on three publicly available Twitter datasets for COVID-19 vaccine stance detection~\cite{glandt2021stance,cotfas2021longest,jiang2022covaxnet}. The final downstream task is designed to be a binary classification, i.e., infer the stance of COVID-vaccine-related tweets as \textit{pro-vaccine} (+1) or \textit{anti-vaccine} (-1). Our final model achieves a high F1 score and accuracy, yielding respectively 0.833 and 0.845. To further evaluate our dataset's final model, we adopt a set of COVID-vaccine-stance-related keywords from \textit{CoVaxxy}~\cite{deverna2021covaxxy} to extract relevant tweets. Then we manually check the machine-generated stance labels. Specifically, we randomly select 500 COVID-vaccine-related tweets (250 \textit{pro-vaccine} and 250 \textit{anti-vaccine}) from our dataset and manually annotate them. Algorithmic and manual annotations have a ``almost perfect'' agreement with Cohen's Kappa Score $k$ = 0.84. This suggests that the final model is capable and reliable in labeling the rest COVID-vaccine-related tweets in our dataset.

\subsection{Data Selection}
To compose a high-quality dataset for the experiment, we apply filters to obtain a set of COVID-vaccine-related PBN with \textit{extreme political bias} and relatively \textit{high social media engagement} from 2021/06 to 2022/06. Particularly, we exclude PBN that is (\textit{i}) unrelated to COVID-19 vaccines; (\textit{ii}) from lean-left and lean-right news media; and (\textit{iii}) with fewer than 100 related social engagements (retweets and replies).


As bot accounts are active on Twitter, we apply Botometer~\cite{yang2022botometer} to filter out malicious bots in our dataset. After that, we apply additional filters to exclude users (\textit{i}) whose locations are outside the United States; and (\textit{ii}) who do not have at least one COVID-vaccine-related tweet before and after consuming COVID-vaccine-related PBN in seven days. After doing so, the subset contains 250 news triplets and 89,535,833 tweets from 17,643 unique users.

\section{RQ1. Behavior Analysis of Three Vaccine Stance Groups}
People with different vaccine stances can have different news consumption behaviors, e.g., pro-vaxxers read more news from the left-leaning media. How they react to a PBN reading intervention may also differ significantly. Therefore, it is critical to identify groups with different vaccine stances to mitigate potential selection and sampling bias. In this section, we (\textit{i}) categorize users into anti-vaccine, pro-vaccine, and vaccine-hesitant groups and then (\textit{ii}) compare their PBN consumption behaviors and social media discussions.

\paragraph{Vaccine Stance Groups.}
Based on a study from the SAGE working group \cite{macdonald2015vaccine}, \textit{vaccine hesitancy} occurs on the continuum between two extremes, i.e., completely accepting (pro-vaccine) and refusing all vaccines (anti-vaccine). Intuitively, vaccine-hesitant users change their stances more frequently compared to anti-vaxxers and pro-vaxxers. Therefore, we first calculate the monthly COVID-vaccine stance changes:
\begin{equation} \label{eq:stance}
    \text{$stance$} = \frac{{P}_\text{before} - {A}_\text{before}}{{P}_\text{before} + {A}_\text{before}}.
\end{equation}
Note that ${P}_\text{before}$ and ${A}_\text{before}$ denote the number of pro-vaccine and anti-vaccine tweets before a user reads PBN, respectively. Similar to~\cite{mitra2016understanding}, we then determine three types of vaccine stance groups based on the following criteria: 
\begin{itemize}[leftmargin=*]
    \item \textit{pro-vaccine group}, users whose monthly COVID-19 vaccine $stance$ are always greater than 0.8;
    \item \textit{anti-vaccine group}, users whose monthly COVID-19 vaccine $stance$ are always less than -0.8; and
    \item \textit{vaccine-hesitant group}, users whose monthly COVID-19 vaccine $stance$ are changing between 0.8 and -0.8.
\end{itemize}
In total, we identify 2,377 pro-vaccine, 1,238 anti-vaccine, and 10,428 vaccine-hesitant users. 

\begin{figure}[!t]
\centering
\includegraphics[width=0.91\columnwidth]{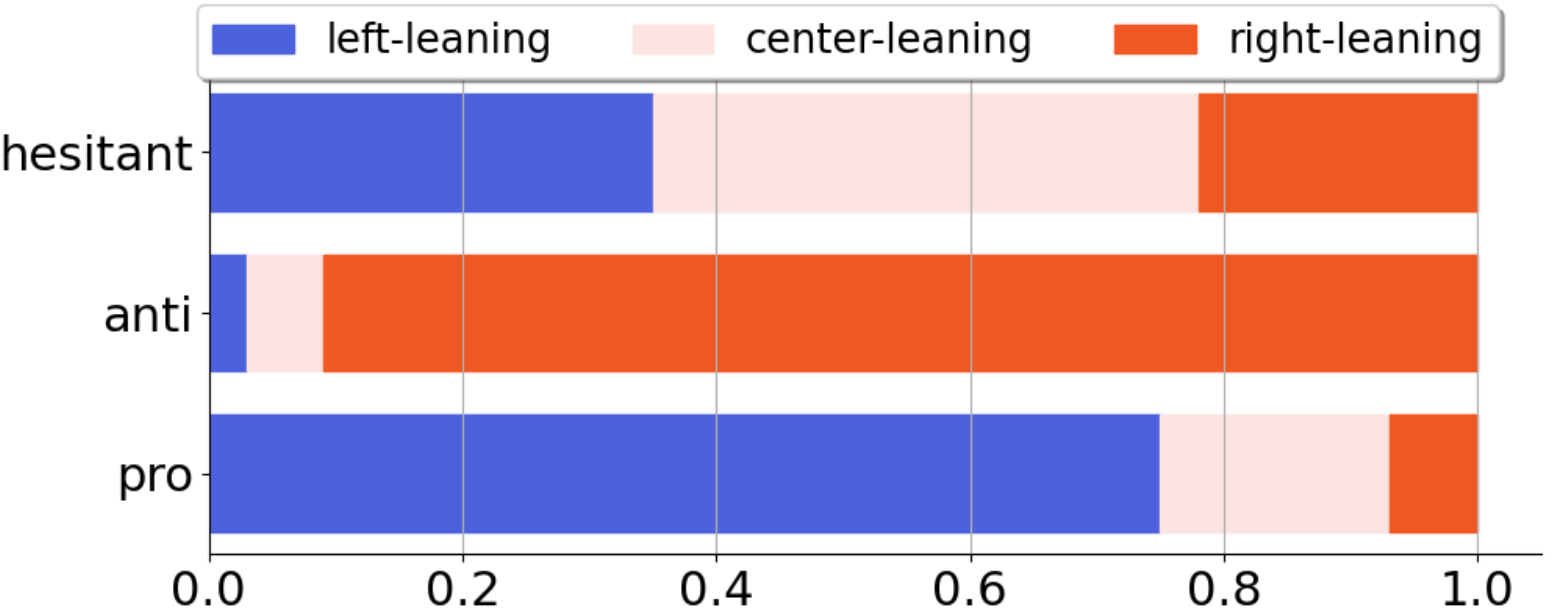} 
\vspace{-.2cm}
\caption{Ratio of left-, right-, and center-leaning news media of pro-vaccine, anti-vaccine, and vaccine-hesitant groups.}
\vspace{-0.3cm}
\label{newsreading}
\end{figure}

\paragraph{News Consumption Behavior.}
Next, we compare the selection bias of three types of news sources in the pro-vaccine group, anti-vaccine group, and vaccine-hesitant group in Figure~\ref{newsreading}. A notable observation is that the pro-vaccine and anti-vaccine groups display a highly \underline{unbalanced distribution} compared to the vaccine-hesitant group. Left-leaning and right-leaning news dominate among the pro-vaccine (around 75\%) and anti-vaccine groups (around 91\%), respectively. Meanwhile, users in the vaccine-hesitant group read news mainly from center-leaning (around 43\%) and left-leaning (around 35\%) media outlets. 

\paragraph{Sociel Media Discussion.}
We further use topic modeling on the collected COVID-vaccine-related tweets to investigate topics each group is interested in after reading PBN. The text is pre-processed by punctuation removal, URL removal, stop words removal, hashtag removal, tokenization, stemming, and lemmatization. The cleaned data is then fed into BERTopic~\cite{grootendorst2022bertopic}, a state-of-the-art topic modeling technique that leverages Sentence-Bert (SBERT)~\cite{reimers2019sentence} for text embedding, UMAP~\cite{mcinnes2018umap} for dimensionality reduction, HDBSCAN~\cite{mcinnes2017hdbscan} for clustering, and a class-based TF-IDF for topic representation. To verify the quality of the result from BERTopic, we manually inspect representative tweets (i.e., tweets nearest to the cluster centroid) for each topic cluster. Finally, we merge over-partitioned clusters to obtain the final list of COVID-related topic clusters.

\begin{table*}[!t]
    \centering
    \resizebox{\linewidth}{!}{\small
    \begin{tabular}{K|j|k}
        \hline
        \textbf{Topic} &  \textbf{Top Words} & \multicolumn{1}{c}{\textbf{Representative Tweets}} \\
        \hline
        Vaccine Refusal & die, kill, serious, allergic, side, effect, risk, freedom & I won’t EVER comply.  \#NOmasks \#NOVaccine. I will deal only in cash, and will only do business with like-minded patriots ...\\ 
        \hline
        Vaccine Acceptance & boost, cdc, fight, child, protect, strong, together, immunity & I am thankful to be fully vaccinated, as earlier this month I recovered from COVID, having had only mild symptoms. I encourage everyone ...\\ 
        \hline
        Conspiracy Theory & fauci, bill, gates, chip, track, lie, kill, bio, weapon, lab, leak & BILL GATES: NOT a doctor. NOT a Scientist. IS a College Dropout. IS a Eugenicist. KNOWN FOR making a computer system susceptible ...\\
        \hline
        Scientific Argument & cell, mrna, evidence, study, immunity, doctor, symptom & According to a study published in Lancet, a single dose of Pfizer or AstraZeneca Covid vaccine offers around 60\% protection against ...\\ 
        \hline        
        Political Narrative & trump, biden, plan, fund, mayor, congress, campaign & The PA legislature has 7 billion from Biden's American Rescue Plan to help PA and they're doing anything else. Political theater on COVID. Not ...\\
        \hline 
    \end{tabular}}
    \vspace{-.2cm}
    \caption{Five topic clusters along with top words (highest TF-IDF scores) and representative tweets (closest to cluster centroids).}
    \vspace{-0.3cm}

    \label{tab:topic}
\end{table*}

\begin{figure}[!t]
    \centering    
        \begin{subfigure}{0.49\textwidth}
        \includegraphics[width=0.91\columnwidth]{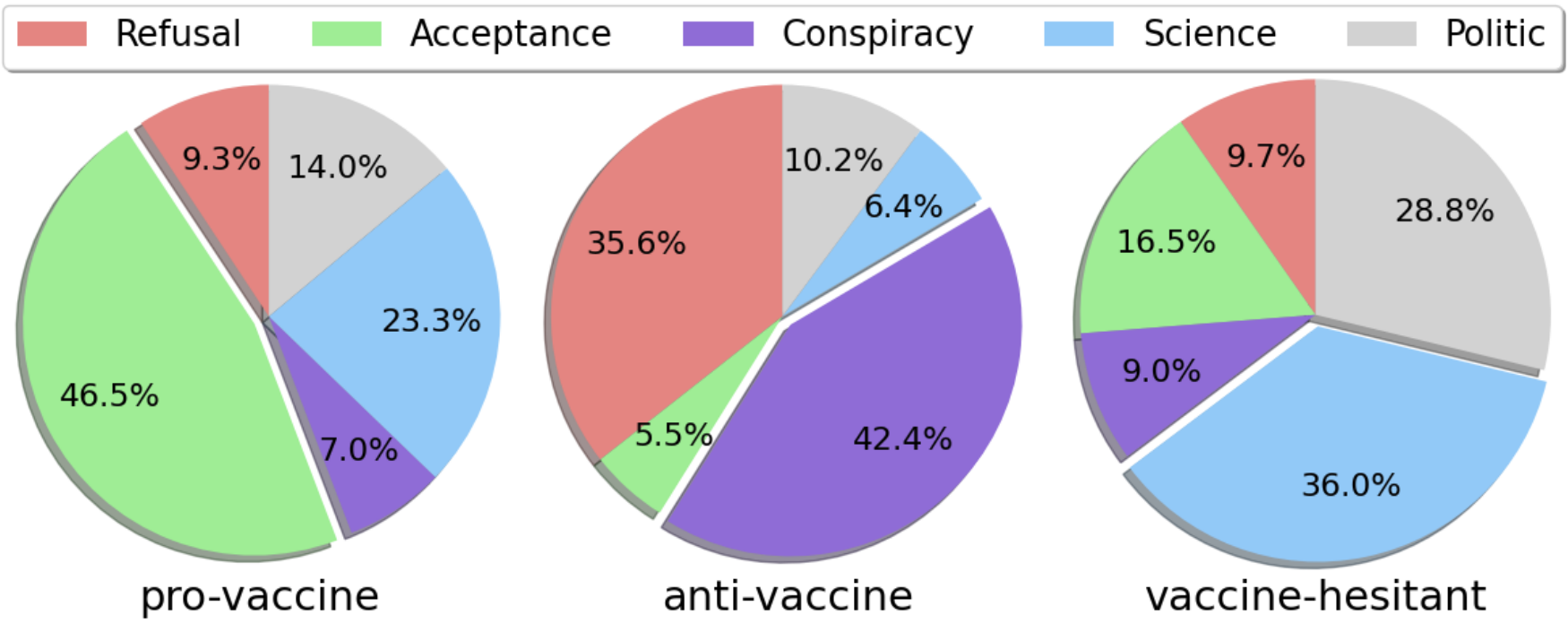} 
        \caption{Overall percentage.}
        \label{piechart}        
        \end{subfigure}

        \begin{subfigure}{0.49\textwidth}
        \includegraphics[width=0.91\columnwidth]{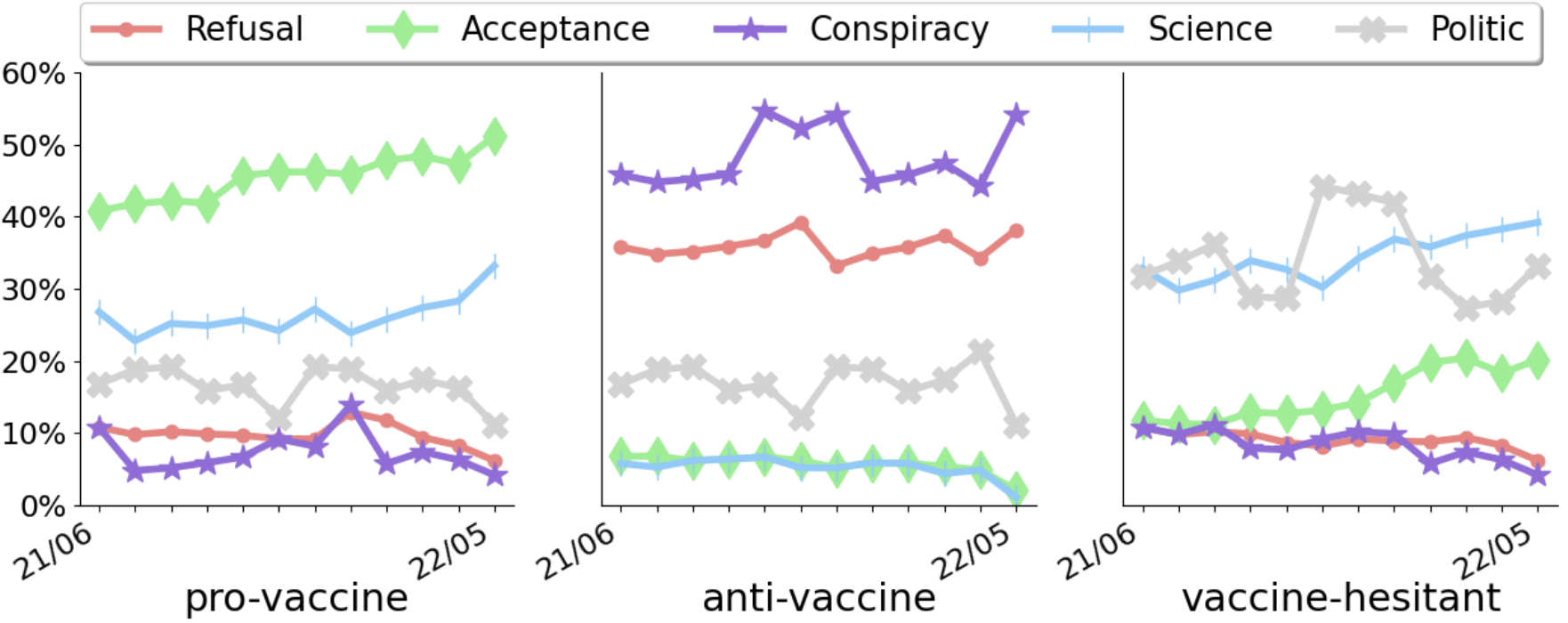} 
        \caption{Monthly percentage.}
        \label{linechart}        
        \end{subfigure}
    \vspace{-.5cm}
    \caption{Overall (a) and monthly (b) percentage of Twitter discussions associated with five COVID-vaccine-related topics among the pro-vaccine group, anti-vaccine group, and vaccine-hesitant group.}
    \vspace{-0.3cm}
    \label{plchart}
\end{figure}

In Table~\ref{tab:topic}, we identify five major COVID-vaccine-related topics: ``vaccine refusal'', ``vaccine acceptance'', \text{``conspiracy theory''}, ``scientific fact'', and ``political narrative''. We illustrate a large variation in the results across these topics in Figure~\ref{piechart}: 

\begin{itemize}[leftmargin=*]
    \item The pro-vaccine group prefers discussions about ``vaccine acceptance'' (46.5\%) and ``scientific argument'' (23.3\%). 
    \item The anti-vaccine group tends to post ``conspiracy theories'' (42.4\%) and ``vaccine refusal'' (35.6\%). 
    \item The vaccine-hesitant group tends to discuss ``scientific arguments'' (36.0\%) and ``politics'' (28.8\%).
\end{itemize}
Figure~\ref{linechart} shows how the discussion changes over time in each stance group. We find that:

\begin{itemize}[leftmargin=*]
    \item ``Vaccine acceptance'' (green) and ``scientific argument'' (blue) discussions have grown steadily in the pro-vaccine and vaccine-hesitant groups.  
    \item Percentages of tweets about ``conspiracy theories'' (purple) and ``vaccine refusal'' (red) show a rapid increase between 2022/04 and 2022/05 in the anti-vaccine group. 
    \item The trending around ``political discussion'' (grey) of the three groups shows a similar pattern.
\end{itemize}

\section{RQ2. Estimating Causal Effect of Reading PBN on Stance Change}  
\begin{figure}[t!]
    \centering
    \begin{subfigure}{0.2\textwidth}
        \includegraphics[width=0.94\columnwidth]{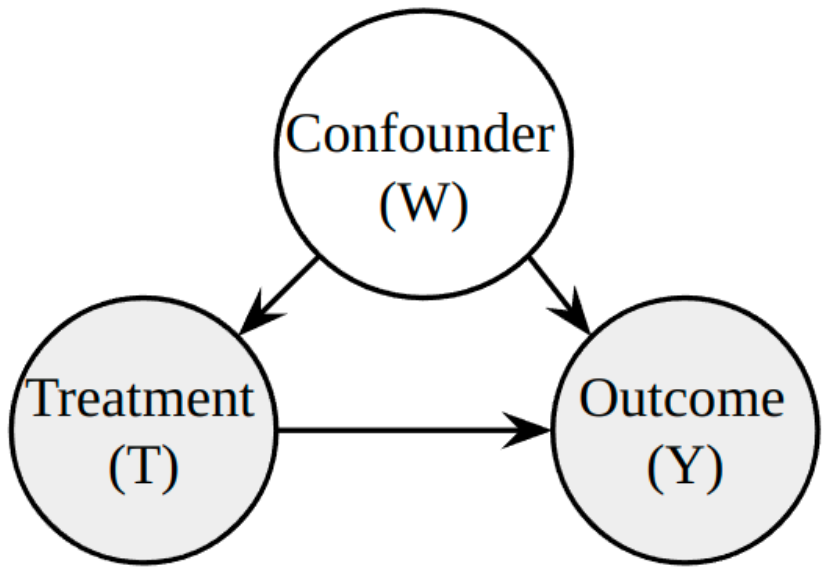}
        \caption{}
        \label{fig:firstdag}
    \end{subfigure}
    \begin{subfigure}{0.26\textwidth}
        \includegraphics[width=0.94\columnwidth]{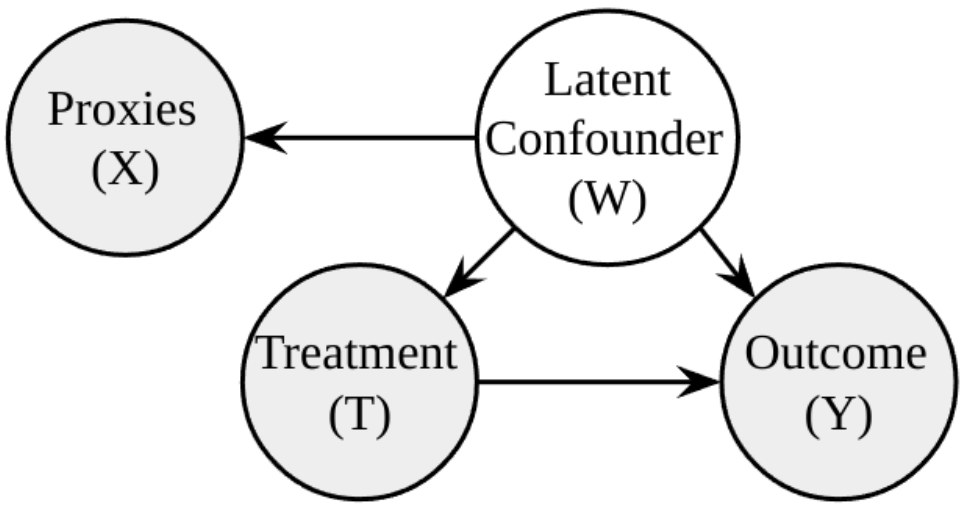}
        \caption{}
        \label{fig:seconddag}
    \end{subfigure}
    \vspace{-.3cm}
    \caption{Two causal DAG of our studied problem. The left one (a) assumes all confounding variables are observed. The right one (b) uses proxies to approximate latent confounders.}
    \label{fig:dag}
    \vspace{-.2cm}
\end{figure}

One of the primary challenges in causal effect estimation with observational data is controlling for confounders. In this research question, we (\textit{i}) identify two types of social media features as \underline{potential confounders}: users' statistical profile feature (e.g., number of tweets) and textual content feature (e.g., historical tweets)~\cite{veitch2020adapting,cheng2022effects}. With the identified potential confounders, we (\textit{i}) estimate the causal effect of reading COVID-vaccine-related PBN on Twitter users’ vaccine stance changes. We hypothesize that reading COVID-vaccine-related news from left-leaning and right-leaning media will \textit{cause} people's COVID-19 vaccine stance to shift toward pro-vaccine and anti-vaccine, respectively. To validate this hypothesis, we begin by formulating the problem as a causal effect estimation task. We use \underline{causal graphs} ~\cite{pearl2009causality} to represent the two scenarios we consider -- (1) when the causal effect is confounded only by observed variables (see Figure~\ref{fig:firstdag}); and (2) when there exist unobserved confounders (see Figure~\ref{fig:seconddag}). With them, we apply state-of-the-art causal learning methods to quantitatively estimate the causal effect using data from \textbf{CovNS}. We also compare the results obtained by the causal learning models and that from a correlation-based model. 

\paragraph{Potential Confounders.}
We consider the following profile features as potential confounders:
\begin{itemize}[leftmargin=*]
    \item The log-transformed number of historical \texttt{tweets}, \texttt{likes}, \texttt{followers}, and \texttt{friends};
    \item The 2020 U.S. presidential election results of the \texttt{location}, ($0=$ blue state, $1=$ red state);
    \item The \texttt{age} (months) of the account;
    \item The Twitter account is \texttt{verified} or not, ($0=$ unverified, $1=$ verified); and
    \item The \texttt{proportion} of COVID-vaccine-related tweets, i.e., a continuum between 0 and 1.
\end{itemize}
For the textual feature, we extract the text embedding of the most recent one-week \texttt{historical tweets} before a user reads PBN from the fine-tuned CT-BERT. These statistical and textual features can reflect user characteristics, which are related to users' online behaviors. Therefore, we select these important social media features as potential confounders for discovering the causal relation in the next phase.

\paragraph{Modeling Causal Relations.}
We define the causal effect we aim to estimate -- reading COVID-vaccine-related PBN (\textit{treatment}) denoted by $T$ on one's COVID-vaccine stance shifting (\textit{outcome}) denoted by $Y$. We represent the causal relations among variables with two different causal graphs~\cite{pearl2009causality} to consider two possible scenarios (see Figure~\ref{fig:dag}).
We consider four different settings in terms of how the values of $T$ and $Y$ are defined (see Table~\ref{tab:causal_setting}).
Then, with $do$-calculus~\cite{pearl2009causality}, we define the causal estimand, i.e., the average treatment effect (ATE):
\begin{equation}
   ATE = E[Y|do(T=1)]-E[Y|do(T=0)],
\end{equation}
where $E[Y|do(T=t)]$ is the expectation of $Y$ when $T$ is intervened to take value $t$.
As aforementioned, we consider the user characteristics, i.e., a set of selected profile features and user history, are related to one's PBN consumption behavior ($T$) and change in COVID-19 vaccine stance ($Y$).
%

In scenario 1, we let the confounder $W$ be these user characteristics. Thus, by conditioning on them, we can block the backdoor path (treatment-confounder-outcome) to handle confounding bias. 
With Figure~\ref{fig:firstdag}, we assume that there is no backdoor path between $T$ and $Y$ by conditioning on the observed user characteristics $W$. This leads to the identification of ATE through backdoor criterion~\cite{pearl2009causality}:
\begin{equation}
    P(Y|do(T)) = \int P(Y|T,W)P(W)dW.
\end{equation}
In this scenario, we estimate ATE with two state-of-the-art causal learning methods. \textit{Double machine learning (DML)}~\cite{chernozhukov2018double} -- \text{DML} estimates heterogeneous treatment effects from observational data with machine learning algorithms. It contains two predictive tasks (i.e., predict the treatment and outcome from the confounder) to ensure unbiased estimates of the causal effect. We use a linear DML in this study. 
\textit{Causal Forest (CF)}~\cite{athey2019generalized} -- CF is widely adopted for causal effect estimation, which performs recursive partition in the confounder space s.t. each leaf of a tree in CF corresponds to a homogeneous subpopulation with similar causal effect.
Compared to linear DML, CF models the relationship between the confounder and the treatment with a tree-based model and infers the causal effect of a test instance by looking up the treatment effect of the subpopulation this instance is mapped to.

Scenario 1 relies on a strong causal assumption that all confounders are observable/measurable (i.e., the unconfoundedness assumption \cite{pearl2009causality}). However, in practice, some confounders are hidden or unmeasurable. For example, one's education and occupation may not be explicitly stated in the user profile. We cannot hope to measure all possible confounders. A common practice is to adopt the ``proxy variables'' \cite{cheng2022estimating}. For example, some textual clues from historical tweets can implicitly reflect users' education levels and jobs. Therefore, in scenario 2, we further relax the unconfoundedness assumption and consider the observed user characteristics $X$ as the \textit{proxies} of the latent confounder $W$. The causal graph is illustrated in Figure~\ref{fig:seconddag}. Then we can leverage proximal causal inference methods~\cite{miao2018identifying} to identify ATE.
Specifically, we consider \textit{Causal effect variational autoencoder
(CEVAE)}~\cite{louizos2017causal} as the estimator. CEVAE leverages deep variational autoencoders (VAE) to learn the representation of the latent confounder given the observed proxy variables.
 
For comparison, we include a correlation-based method. Specifically, we implement a simple \textit{logistic regression on treatment} that predicts the outcome $Y$ with the treatment $T$ as its input. It can be considered as a generalized version of the naive estimator~\cite{rubin1978bayesian} that does not control the effect of confounding variables.

\begin{figure*}[!t]
    \centering \small
    \begin{subfigure}{0.49\textwidth}
        \includegraphics[width=1\columnwidth]{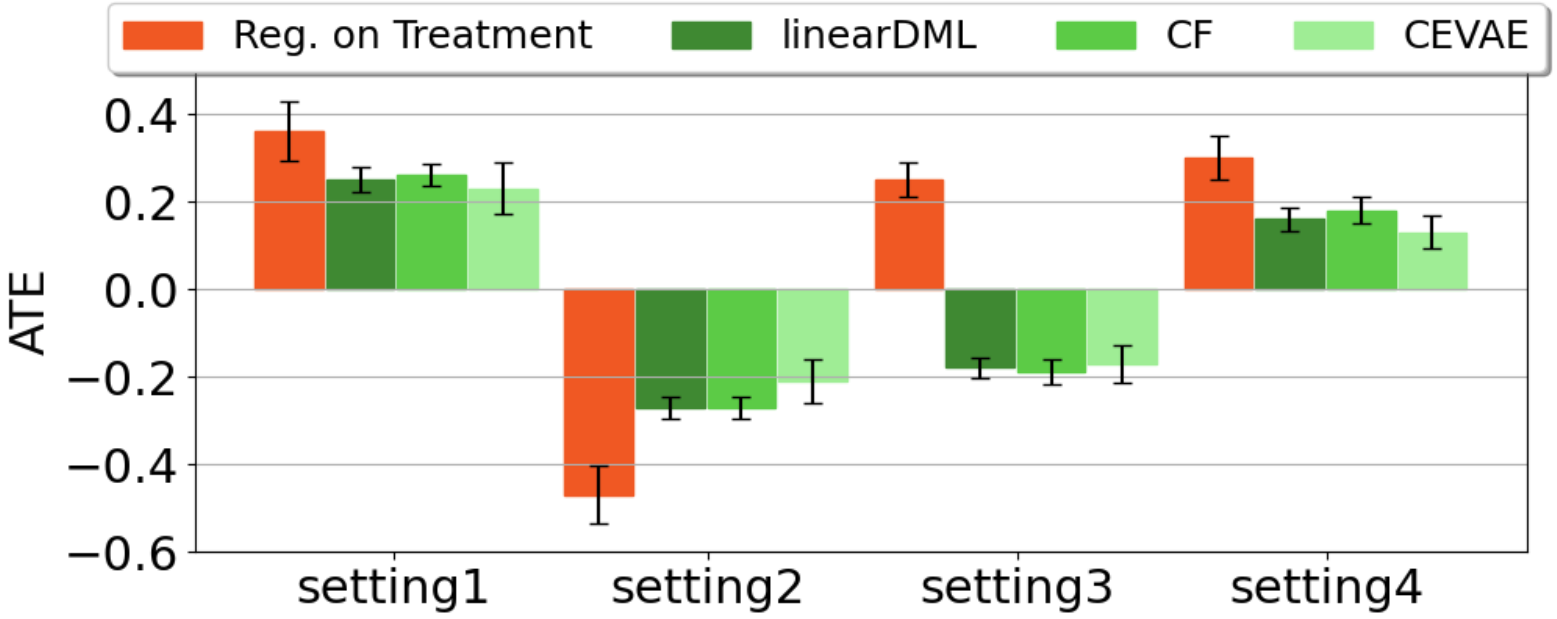}
        \caption{Overall.}
        \label{fig:allce}
    \end{subfigure}
    \begin{subfigure}{0.49\textwidth}
        \includegraphics[width=1\columnwidth]{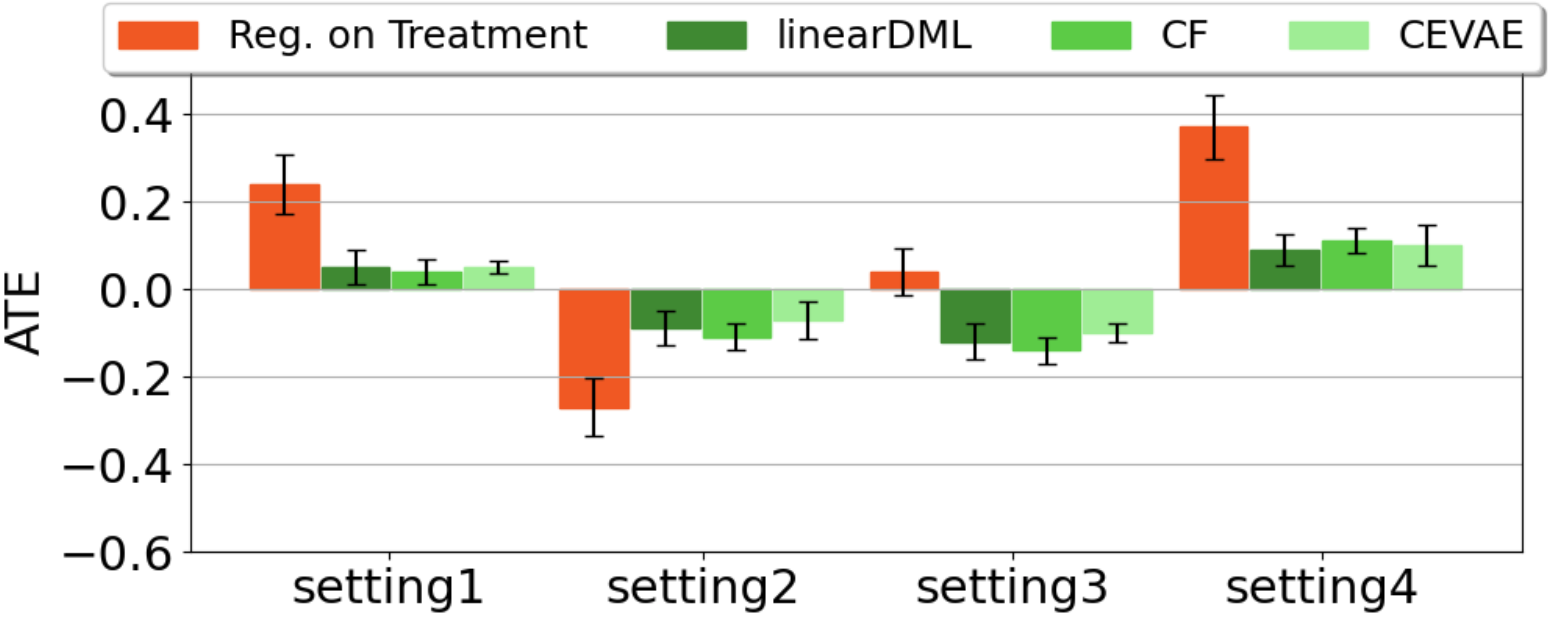}
        \caption{Pro-vaccine group.}
        \label{fig:firstce}
    \end{subfigure}
    \begin{subfigure}{0.49\textwidth}
        \includegraphics[width=1\columnwidth]{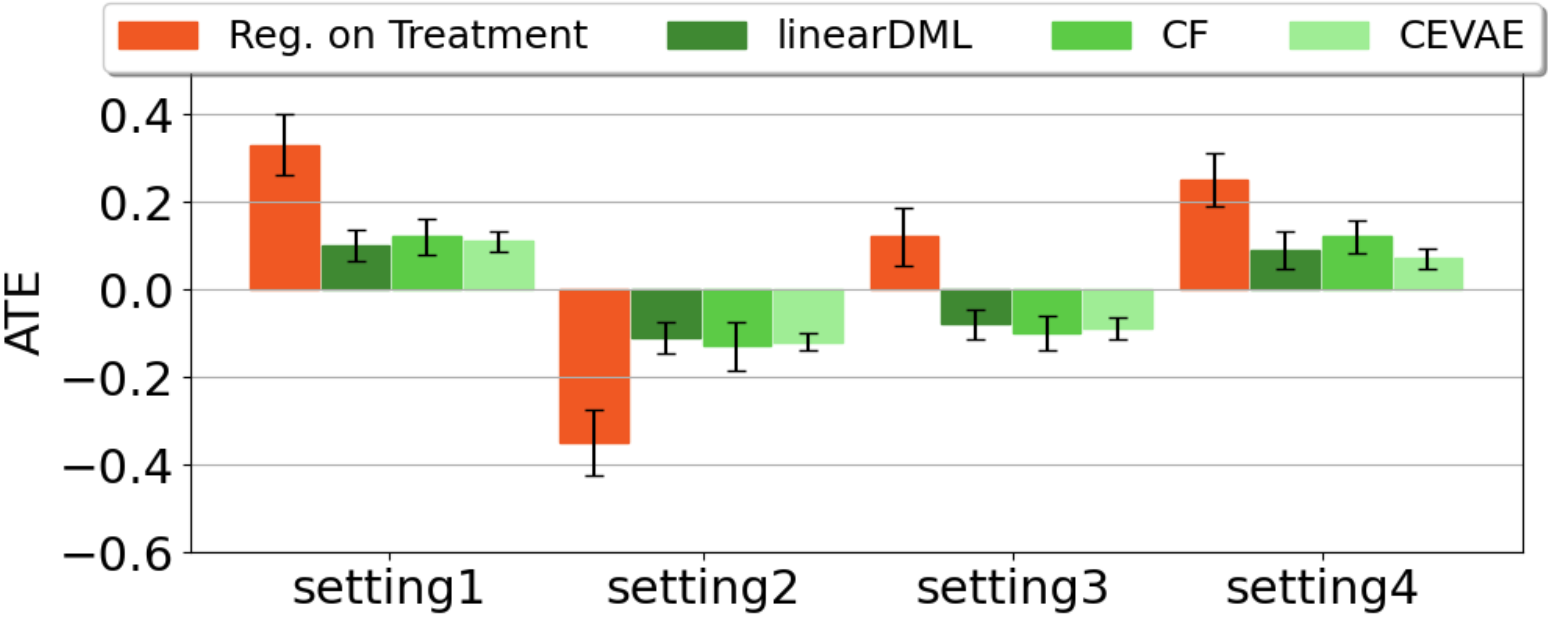}
        \caption{Anti-vaccine group.}
        \label{fig:thirdce}
    \end{subfigure}
    \begin{subfigure}{0.49\textwidth}
        \includegraphics[width=1\columnwidth]{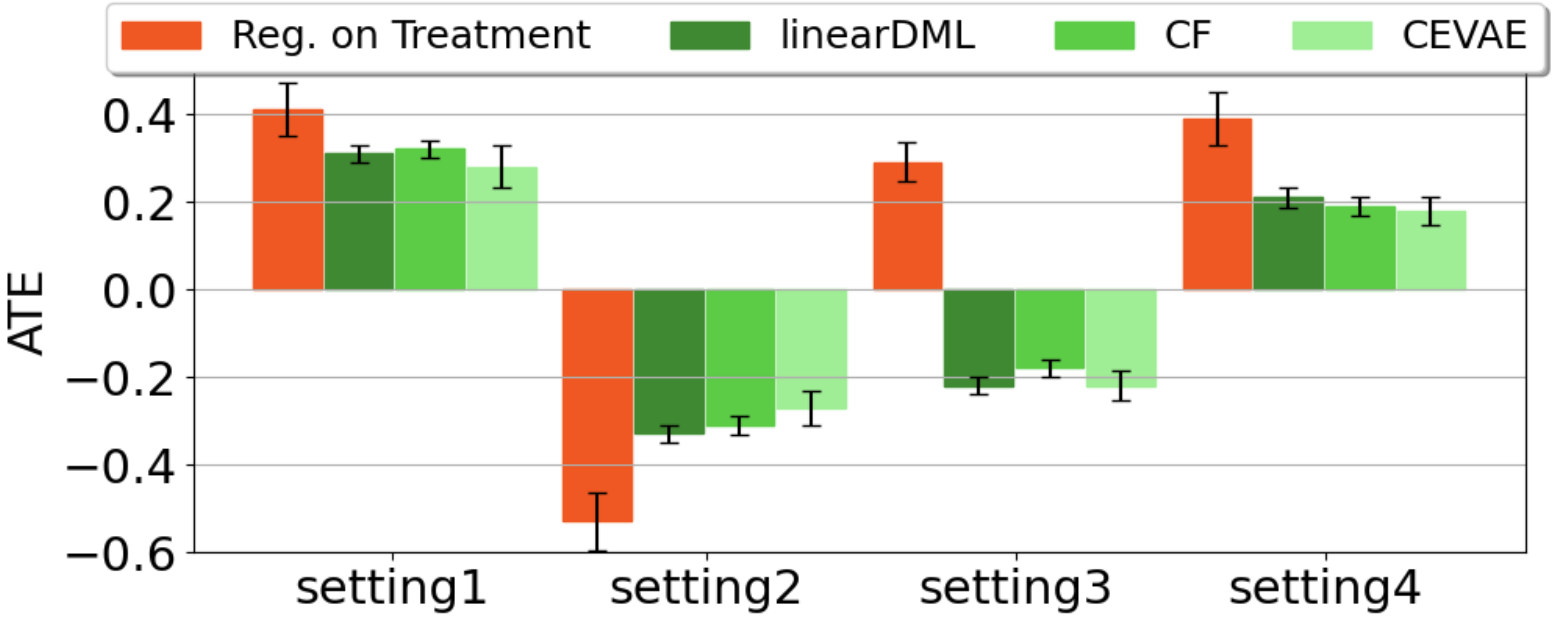}
        \caption{Vaccine-hesitant group.}
        \label{fig:secondce}
    \end{subfigure}
    \vspace{-.2cm}
        \caption{The overall estimated ATE and that of each (pre-treatment) stance group: we compare correlation-based (red) and causal learning models (green) in $4$ experimental settings. We ran each model $10$ times to report the mean values (histograms) and standard deviations (error bars). Note that the ATE inferred by the causal learning methods are similar ($p > 0.05$ for t-test) while they are significantly different from the one estimated by the correlation-based model ($p < 0.001$ for t-test).}
    \vspace{-0.2cm}
    \label{fig:ce4}
\end{figure*}

\begin{table}[!t]
    \centering
    \resizebox{\linewidth}{!}{\small
    \begin{tabular}{cll}
        \toprule
        \textbf{Setting} & \multicolumn{1}{c}{\textbf{Treatment (T)}} & \multicolumn{1}{c}{\textbf{Outcome (Y)}} \\
        \midrule
        \multirow{2}{*}{1} & {0: read \textit{center} news} &  {0: unchanged stance} \\
        & {1: read \textit{left} news} &  {1: move toward pro-vax} \\
        \midrule
        \multirow{2}{*}{2} & {0: read \textit{center} news} &  {0: unchanged stance} \\
        &{1: read \textit{left} news} &  {1: move toward anti-vax} \\
        \midrule
        \multirow{2}{*}{3} & {0: read \textit{center} news} &  {0: unchanged stance} \\
        {} & {1: read \textit{right} news} &  {1: move toward pro-vax} \\
        \midrule
        \multirow{2}{*}{4} & {0: read \textit{center} news} &  {0: unchanged stance} \\
        {} &{1: read \textit{right} news} &  {1: move toward anti-vax} \\
        \bottomrule
    \end{tabular}}
    \vspace{-.3cm}    
    \caption{Experiment settings for the causal study.}
    \vspace{-.3cm}    
    \label{tab:causal_setting}
\end{table}

\paragraph{Estimating Causal Effects.} Figure~\ref{fig:ce4} shows the comparison of the overall estimated ATE by different methods and that of the pro-vaccine group (\ref{fig:firstce}), anti-vaccine group (\ref{fig:thirdce}), and vaccine-hesitant group (\ref{fig:secondce}). The estimated ATE from \text{DML}, \text{CF}, and \text{CEVAE} are similar ($p>0.05$).

\begin{itemize}[leftmargin=*]
    \item Comparing the results from settings 1 and 2, we observe that reading PBN from left-leaning media causes individuals to be \underline{more pro-vaccine}. In contrast, results from settings 3 and 4 show that consuming PBN from right-leaning media causes people to be \underline{more anti-vaccine}. 
    \item Results from logistic regression (LR) on treatment are outliers in all experiments ($p<0.001$). For example, the estimated ATEs of LR in setting 3 show that reading right-leaning PBN will make people more pro-vaccine, which is opposite to the results from causal learning models. A possible explanation is that the results of correlation-based observational studies usually suffer from spurious correlation due to \underline{confounding bias}, especially when the confounding effect is important~\cite{austin2011introduction}. 
    \item In addition, we find that the magnitudes of the estimated ATE of the vaccine-hesitant group are higher than the pro-vaccine group and anti-vaccine group, indicating that vaccine-hesitant people are more likely to \underline{change their vaccine stances} after consuming PBN. 
\end{itemize}





\section{Implications}
\textbf{RQ1.} \textit{Users with varying stances on vaccines display unique patterns of PBN consumption and social media discussion. Future research should consider the disparity when studying vaccine-related problems in social media.}

\begin{itemize}[leftmargin=*]
    \item Compared to the vaccine-hesitant people, the pro-vaccine and anti-vaccine group members are more likely to read PBN from \underline{far-left} and \underline{far-right} news media, respectively. Intensive exposure to highly biased news may explain the occurrence of extreme COVID-19 vaccine stances.
    \item As ``scientific arguments'' and ``political narratives'' usually contain debatable opinions, the vaccine-hesitant group member is able to gather \underline{diverse information} from both sides. Thus, their stances may move back and forth.
    \item The pattern of increasing interest in ``vaccine acceptance'' and a decline in ``vaccine refusal'' indicates that the vaccine-hesitant group is becoming \underline{more positive} about the vaccine.
\end{itemize}

\noindent \textbf{RQ2.} \textit{The exposure to COVID-vaccine-related news from left- and right-leaning media \textit{causes} one's COVID-19 vaccine stance shift toward pro- and anti-vaccine respectively, especially for users in the vaccine-hesitant group. This suggests that it is possible for malicious people to {manipulate public opinion} through PBN interventions.}

\begin{itemize}[leftmargin=*]
    \item The similar estimated causal effects from causal learning models indicate that the unobserved confounding variables may have limited impacts on the treatment and outcome.
    \item Comparing the magnitudes of the estimated ATE (see Figure~\ref{fig:allce}), we observe that left-leaning news (setting 1 and 2) is more \textit{influential} than right-leaning news (setting 3 and 4). We speculate that vaccine-hesitant people are more likely to \underline{become pro-vaxxers} than anti-vaxxers over time through the influence of PBN.    
    \item As reading left-leaning news has a very small causal effect (around $0.1$) on anti-vaxxers' vaccine stance changes (see Figure~\ref{fig:thirdce}), it is {unlikely} to alter anti-vaxxers to pro-vaxxers through PBN reading interventions. Meanwhile, pro-vaxxers are difficult to become anti-vaxxers by consuming PBN (see Figure~\ref{fig:firstce}). This indicates that people with extreme views about vaccines tend to \underline{reinforce} their existing stances. On the other hand, the vaccine-hesitant majority are vulnerable to the influence of PBN and, thus, may frequently \underline{reverse} their minds. 
\end{itemize}

\section{Conclusions}
This paper investigates the impact of PBN consumption on the vaccine stance changes of social media users. We construct \textbf{CovNS}, which contains data from both news media and social media. We compare the PBN consumption behavior and social media discussion between three vaccine stance groups. By identifying potential confounders, we leverage state-of-the-art causal inference methods to estimate the causal effect. Our experiments and analyses have implications for fostering the research of vaccine hesitancy in social media. We conclude that consuming left-leaning and right-leaning news causes people to be pro-vaccine and anti-vaccine, respectively. More importantly, there is only a small possibility for anti-vaxxers to become pro-vaxxers via PBN reading interventions on social media, and vice versa.

\bibliographystyle{named}
\bibliography{ijcai24}

\end{document}